\documentclass[aps,prl,twocolumn,showpacs,floatfix]{revtex4}
\usepackage{bbm,amsmath,amssymb,amsbsy,graphicx,times,subfigure}

\usepackage[latin1]{inputenc}
\renewcommand{\det}{{\rm Det}\,}
\newcommand{\gr}[1]{\boldsymbol{#1}}
\newcommand{\be}{\begin{equation}}
\newcommand{\ee}{\end{equation}}
\newcommand{\bea}{\begin{eqnarray}}
\newcommand{\eea}{\end{eqnarray}}

\newcommand{\eq}[1]{Eq.~(\ref{#1})}

\begin{document}
\title{Entanglement in Gaussian matrix-product states}
\date{August 19, 2006}
\author{Gerardo Adesso$^{1,2}$ and Marie Ericsson$^1$}

\affiliation{$^1$Centre for Quantum Computation, DAMTP, Centre for
Mathematical Sciences, University of Cambridge, Wilberforce Road,
Cambridge CB3 0WA, United Kingdom \\ $^2$Dipartimento di Fisica ``E.
R. Caianiello'', Universit\`a degli Studi di Salerno, INFN Sezione
di Napoli-Gruppo Collegato di Salerno, Via S. Allende, 84081
Baronissi (SA), Italy}


\begin{abstract}
%We study Gaussian matrix product states,
% of continuous variable systems,
%introduced in Ref. \cite{GMPS} as states of harmonic lattices,
%constructed by projections from an ancillary space of
%infinitely-entangled bonds connecting neighbor sites. We describe
%the projection operation applied at each site via a canonically
%associated Gaussian state, the \emph{building block}. We analyze in
%detail the distribution of continuous-variable entanglement in
%Gaussian matrix product states, and demonstrate that their quantum
%correlation length is determined by the degree of entanglement in
%the building block. In particular, an infinitely entangled building
%block leads to fully permutation-invariant Gaussian states, whose
%properties are interpreted within the matrix product framework. This
%reveals a basic difference compared to the case of spin systems:
%Gaussian matrix product states can possess unlimited, long-range
%entanglement even with minimum number of ancillary bonds. Finally we
%show how these states can be experimentally engineered in a
%realistic setting from $N$ copies of a three-mode building block and
%$N$ two-mode finitely squeezed states.
Gaussian matrix product states are obtained as the outputs of
projection operations from an ancillary space of $M$ infinitely
entangled bonds connecting neighboring sites, applied at each of
$N$ sites of an harmonic chain. Replacing the projections by
associated Gaussian states, the {\em building blocks}, we show
that the entanglement range in translationally-invariant Gaussian
matrix product states depends on how entangled the building blocks
are. In particular, infinite entanglement in the building blocks
produces fully symmetric
 Gaussian states with maximum entanglement
range. From their peculiar properties of entanglement sharing, a
basic difference with spin chains is revealed: Gaussian matrix
product states can possess unlimited, long-range entanglement even
with minimum number of ancillary bonds ($M$=1). Finally we discuss
how these states can be experimentally engineered from $N$ copies
of a three-mode building block and $N$ two-mode finitely squeezed
states.
\end{abstract}
\maketitle

\noindent {\em Introduction.}--- The description of
 many-body systems and the understanding of multiparticle
 entanglement are among the hardest challenges of quantum physics.
 The two issues are entwined: recently, the basic tools
 of quantum information theory have found useful applications
 in condensed matter physics. In particular, the formalism of \emph{matrix product
 states} (MPS) \cite{maria} has led to an efficient simulation
 of many-body spin Hamiltonians \cite{vidal2} and to a deeper understanding of quantum phase transitions \cite{wolfito}.

Beyond qubits, and discrete-variable systems in general, a growing
interest is being witnessed in the theoretical and experimental
applications of so-called continuous variable (CV) systems, such as
ultracold atoms or light modes, to quantum information and
communication processings \cite{review}. Besides their usefulness in
feasible implementations, quasi-free states of harmonic lattices,
best known as Gaussian states, are endowed with structural
properties that make the characterization of their entanglement
amenable to an analytical analysis \cite{adebook}. Within this
context, the extension of the matrix product framework to Gaussian
states of CV systems has been recently introduced as a possible tool
to prove an entropic area law for critical bosonic systems on
harmonic lattices \cite{GMPS}, thus complementing the known results
for the non-critical case \cite{area}.

In this work we adopt a novel point of view, aimed to comprehend
the correlation picture of the considered many-body systems from
the physical structure of the underlying MPS framework. In the
case of harmonic lattices, we demonstrate that  the quantum
correlation length (the maximum distance between pairwise
entangled sites) of translationally invariant Gaussian MPS is
determined by the amount of entanglement encoded in a smaller
structure, the `building block', which is a Gaussian state
isomorphic to the MPS projector at each site. This connection
provides a series of necessary and sufficient conditions for
bipartite entanglement of distant pair of modes in Gaussian MPS
depending on the parameters of the building block, as explicitly
shown for a six-mode harmonic ring.
%This connection, analytically shown for some explicit instances,
%is evidenced in the Jamiolkowski picture of Gaussian operations
%\cite{giedke}, and enables us to
%, and to the best of our knowledge has never been
%pursued in finite-dimensional MPS.
%shed new light on the physical relationship between the form of MPS
%projectors and the properties of the output state.
For any size of the ring we show remarkably that, when single
ancillary bonds connect neighboring sites, an infinite
entanglement in the building block leads to fully symmetric
(permutation-invariant) Gaussian MPS where each individual mode is
equally entangled with any other, independently of the distance.
As the block entropy of these states can diverge for any
bipartition of the ring \cite{adescaling}, our results unveal a
basic difference with finite-dimensional MPS, whose entanglement
is limited by the bond dimensionality \cite{vidal} and is
typically short-ranged \cite{kore}. Finally, we demonstrate how to
experimentally implement the MPS construction to produce multimode
Gaussian states useful for CV communication networks
\cite{network}.
% starting from many copies of entangled
%states of few modes.

\smallskip

\noindent {\em Gaussian matrix product states.}--- In a CV system
consisting of $N$ canonical bosonic modes,  described by the vector
$\hat{R} = \{\hat q_1, \hat q_2, \ldots, \hat q_N, \hat p_1, \hat
p_2 \ldots, \hat p_N\}$ of the field quadrature operators, Gaussian
states (such as coherent and squeezed states) are fully
characterized by the first statistical moments (arbitrarily
adjustable by local unitaries: we will set them to zero) and by the
$2N \times 2N$ real symmetric covariance matrix (CM) $\gr\gamma$ of
the second moments
$\gamma_{ij}=1/2\langle\{\hat{R}_i,\hat{R}_j\}\rangle$
\cite{adebook}.

The Gaussian matrix product states introduced in Ref.~\cite{GMPS}
are $N$-mode states obtained by taking a fixed number, $M$, of
infinitely entangled ancillary bonds (EPR pairs) shared by adjacent
sites, and applying an arbitrary $2M \rightarrow 1$ Gaussian
operation on each site $i=1,\ldots,N$. Here the {\em cardinality}
$M$ is the CV counterpart of the dimension $D$ of the matrices in
standard MPS \cite{maria}. Such a process
%, which corresponds to
%construct a state via entanglement swapping by the ancillary
%(valence) bonds,
can be better understood by resorting to the
Jamiolkowski isomorphism between (Gaussian) operations and
(Gaussian) states \cite{giedke}. In this framework, one starts with
a chain of $N$ Gaussian states of $2M+1$ modes: the {\em building
blocks}. The global Gaussian state of the chain is described by a CM
$\gr\Gamma = \bigoplus_{i=1}^N \gr\gamma^{[i]}$. As the interest in
MPS lies mainly in their connections with ground states of
Hamiltonians invariant under translation \cite{GMPS}, we can focus
on pure ($\det \gr\gamma^{[i]} = 1$), translationally invariant
($\gr\gamma^{[i]} \equiv \gr\gamma \, \forall i$) Gaussian MPS.
Moreover, in this work we consider single-bonded MPS, i.e.~with
$M=1$.
%This is also physically motivated in view of experimental
%implementations of Gaussian MPS, as more than one EPR bond would
%result in a building block with five or more correlated modes, which
%appears technologically demanding.
However, our analysis easily
generalizes to multiple bonds, and to mixed Gaussian states as well.

Under the considered  prescriptions, the building block
$\gr\gamma$ is a pure Gaussian state of the three modes with
respective CM $\gr\alpha_{1,2,3}$. As we aim to construct a
translationally invariant state, it is convenient to consider a
$\gr\gamma$ whose first two modes  have the same reduced CM. This
yields a bisymmetric \cite{adescaling}, pure, three-mode Gaussian
building block whose CM $\gr\gamma$ can be written without loss of
generality in standard form \cite{3modi}, with
$(\det\gr\alpha_1)^{1/2} = (\det\gr\alpha_2)^{1/2} \equiv s$ and
$(\det\gr\alpha_3)^{1/2} \equiv x$. This choice of the building
block is physically motivated by the fact that, among all pure
three-mode Gaussian states, bisymmetric states maximize the
genuine tripartite entanglement \cite{3modi}.
%: no entanglement is
%thus wasted in the projection process.
It is instructive to write
$\gr\gamma$ in the block form
\begin{equation}\label{bblock}
\gr\gamma = \left(
           \begin{array}{cc}
             \gr\gamma_{ss} & \gr\gamma_{sx} \\
             \gr\gamma_{sx}^T & \gr\gamma_x \\
           \end{array}
         \right)\!,
\end{equation}
where $\gr\gamma_{ss}$ is the CM of modes $1$ and $2$, $\gr\gamma_x$
is the CM of mode $3$, and the intermodal correlations are encoded
in $\gr\gamma_{sx}$. Explicitly \cite{3modi}:
$\gr\gamma_{ss}={\footnotesize{\left(
                                          \begin{array}{cc}
                                            s & t_+ \\
                                            t_+ & s \\
                                          \end{array}
                                        \right)
\oplus\left(
                                          \begin{array}{cc}
                                            s & t_- \\
                                            t_- & s \\
                                          \end{array}
                                        \right)}}$, with
$t_{\pm}=[x^2-1 \pm \sqrt{16 s^4 - 8 (x^2 + 1) s^2 + (x^2 -
1)^2}]/(4s)$; $\gr\gamma_x = {\rm diag}\{x,\,x\}$; and
$\gr\gamma_{sx}^T ={\footnotesize{ \left(
                                                                 \begin{array}{cccc}
                                                                   u_+ & u_+ & 0 & 0 \\
                                                                   0 & 0 & u_- & u_- \\
                                                                 \end{array}
                                                               \right)}}$,
                                                               with
$u_\pm = \frac{1}{4} \sqrt{\frac{x^2 - 1}{s x}} \left[\sqrt{(x - 2
s)^2 - 1} \pm \sqrt{(x + 2 s)^2 - 1}\right]$.

\begin{figure}[t!]
\includegraphics[width=6cm]{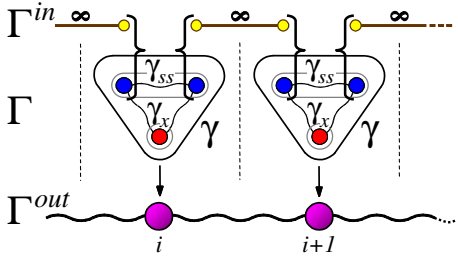} \caption{(Color online) Gaussian matrix product states.
$\gr\Gamma^{in}$ is the state of $N$ EPR bonds and $\gamma$ is the
three-mode building block. After the EPR measurements (depicted as
curly brackets), the modes $\gr\gamma_x$ collapse into a Gaussian
MPS with global state $\gr\Gamma^{out}$.} \label{fiocco}
\end{figure}

The MPS construction works as follows (see Fig.~\ref{fiocco}). The
global CM $\gr\Gamma = \bigoplus_{i=1}^N \gr\gamma$ corresponds to
the projector from the  state $\gr\Gamma^{in}$ of the $N$ ancillary
EPR pairs, to the final $N$-mode Gaussian MPS $\gr\Gamma^{out}$.
This is realized by collapsing the  state $\gr\Gamma^{in}$,
transposed in phase space, with the input port
$\gr\Gamma_{ss}=\bigoplus_i \gr\gamma_{ss}$ of $\gr\Gamma$, so that
the output port $\gr\Gamma_{x} = \bigoplus_i \gr\gamma_x$ turns into
the desired $\gr\Gamma^{out}$. Here collapsing means that, at each
site, the two two-mode states, each constituted by one mode ($1$ or
$2$)  of $\gr\gamma_{ss}$ and one half of the EPR bond between site
$i$ and its neighbor ($i-1$ or $i+1$, respectively), undergo an
``EPR measurement'' i.e.~are projected onto the infinitely entangled
EPR state \cite{giedke,GMPS}. An EPR pair is described by a CM
\begin{equation}\label{tmss}
\gr\sigma_{i,j}(r) = {\scriptsize{ \left(\!\!
  \begin{array}{cc}
    \cosh(2r) & \sinh(2r) \\
    \sinh(2r) & \cosh(2r) \\
  \end{array}
\!\!\right)\! \oplus \!\left(\!\!
  \begin{array}{cc}
    \cosh(2r) & -\sinh(2r) \\
    -\sinh(2r) & \cosh(2r) \\
  \end{array}
\!\!\right)}},
\end{equation}
which corresponds to a two-mode squeezed state of modes $i$ and $j$,
in the limit of infinite squeezing ($r \rightarrow \infty$). The
input state is then $\gr\Gamma^{in} = \lim_{r \rightarrow \infty}
\bigoplus_{i}^{N} \gr\sigma_{i,i+1}(r)$, where we have set periodic
boundary conditions so that $N+1 = 1$ in labeling the sites.
%In  the end, the limit of
%infinite We anticipate that the results derived in the following
%will not strictly rely on the assumption of an infinite squeezing
%(and thus infinite entanglement in the bonds): this condition will
%be in fact relaxed in the end, adding feasibility to our scheme.
The projection corresponds mathematically to taking a Schur
complement (see Refs.~\cite{GMPS,giedke} for details), yielding an
output pure Gaussian MPS of $N$ modes on a ring with a CM
\begin{equation}\label{cmout}
\gr\Gamma^{out} = \gr\Gamma_{x} - \gr\Gamma_{sx}^T (\gr\Gamma_{ss} +
\gr\theta \gr\Gamma^{in} \gr\theta)^{-1} \gr\Gamma_{sx}\,,
\end{equation}
where $\gr\Gamma_{sx}=\bigoplus_i \gr\gamma_{sx}$, and $\gr\theta
=\bigoplus_i {\rm diag}\{1,\,1,\,-1,\,-1\}$ represents transposition
in phase space ($\hat q_i \rightarrow \hat q_i,\,\hat p_i
\rightarrow - \hat p_i$).

First we note that the Gaussian states constructed in this way are
ground states of harmonic Hamiltonians (a property of all Gaussian
MPS \cite{GMPS}). This follows as no mutual correlations are created
between the operators $\hat q_i$ and $\hat p_j$ for any
$i,j=1,\ldots,N$, having  both EPR bonds and building blocks in
standard form. The final CM \eq{cmout} thus takes the form
\begin{equation} \label{circ}
\gr\Gamma^{out} = C^{-1} \oplus C\,, \end{equation} where $C$ is a
circulant $N \times N$ matrix. It can be shown that
 a CM of the form \eq{circ} corresponds to the ground state of the quadratic
Hamiltonian $\hat H = \frac12\big(\sum_{i} \hat p_i^2 + \sum_{i,j}
\hat q_i V_{ij} \hat q_j \big)$, with the potential matrix given by
$V=C^2$ \cite{chain}.

\smallskip

\noindent {\em Entanglement distribution.}--- In the Jamiolkowski
picture \cite{GMPS,giedke}, different MPS projectors correspond to
differently entangled Gaussian building blocks. Let us
 recall that, according to the ``positivity of partial
transposition'' (PPT) criterion, a Gaussian state is separable (with
respect to a $1 \times N$ bipartition) if and only if the partially
transposed CM
%, obtained
% performing trasposition (time reversal in phase space)
% in the subspace of one subsystem only,
satisfies the uncertainty principle \cite{simon}. As a measure of
entanglement, for two-mode symmetric Gaussian states
$\gr\gamma_{i,j}$ the entanglement of formation $E_F$ is computable
via the formula \cite{efprl}: $E_F (\gr\gamma_{i,j}) = \max \{0,\,
f(\eta_{i,j})\}$,
 with $f(x) = \frac{(1+x)^2}{4x}
\log{\frac{(1+x)^2}{4x}} - \frac{(1-x)^2}{4x}
\log{\frac{(1-x)^2}{4x}}$. Here the positive parameter $\eta_{i,j}$
is the smallest symplectic eigenvalue of the partial transpose  of
$\gr\gamma_{i,j}$. For a two-mode state, $\eta_{i,j}$ can be
computed from the symplectic invariants of the state
\cite{extremal}, and the PPT criterion simply yields
$\gr\gamma_{i,j}$ entangled as soon as $\eta_{i,j}<1$, while
infinite entanglement
%(accompanied by infinite energy in the state)
is reached for $\eta_{i,j} \rightarrow 0^+$.

We are interested in studying the quantum correlations of Gaussian
MPS of the form as in \eq{cmout}, and in relating them to the
entanglement properties of the building block $\gr\gamma$. The CM
in \eq{bblock} describes a physical state if $x \ge 1$ and $s\ge
s_{\min}\equiv(x+1)/2$ \cite{3modi}. At fixed $x$, and so at fixed
CM of mode $3$ (output port), the entanglement in the CM
$\gr\gamma_{ss}$ of the first two modes (input port) is
monotonically increasing as a function of $s$ (as it can be
checked by studying the respective symplectic eigenvalue
$\eta_{ss}$), ranging from the case $s=s_{\min}$ when
$\gr\gamma_{ss}$ is separable to the limit $s \rightarrow \infty$
when the block $\gr\gamma_{ss}$ is infinitely entangled.
Accordingly, the entanglement between each of the first two modes
of $\gr\gamma$ and the third one decreases with $s$. The main
question we raise is how the initial entanglement in the building
block $\gr\gamma$ gets distributed in the Gaussian MPS
$\gr\Gamma^{out}$. The answer will be that the more entanglement
we prepare in the input port $\gr\gamma_{ss}$, the longer the
range of the quantum correlations in the output MPS will be. We
start from the case of minimum $s$.

\noindent {\em Short-range correlations.}--- Let us consider a
building block $\gr\gamma$ with $s=s_{\min}=(x+1)/2$.
%This three-mode Gaussian
%state can be produced experimentally and is useful, for instance, to
%perform $1 \rightarrow 2$ telecloning of unknown coherent states
%\cite{telecloning}.
It is straightforward to evaluate, as a function of $x$, the
Gaussian MPS in \eq{cmout} for an arbitrary number of modes (we omit
the CM here, as no particular insight can be drawn from the the
explicit expressions of the covariances). By repeatedly applying the
PPT criterion, one can analytically check that each reduced two-mode
block $\gr\gamma^{out}_{i,j}$ is separable for $|i-j|>1$, which
means that the output MPS $\gr\Gamma^{out}$ exhibits bipartite
entanglement only between nearest neighbor modes, for any value of
$x>1$ (for $x=1$ we  obtain a product state). While this certainly
entails that $\gr\Gamma^{out}$ is genuinely multiparty entangled,
due to the translational invariance, it is interesting to observe
that, without feeding entanglement in the input port
$\gr\gamma_{ss}$ of the original building block, the range of
quantum correlations in the output MPS is  minimum. The pairwise
entanglement between nearest neighbors will naturally decrease with
increasing number of modes, being frustrated by the overall symmetry
and by the intrinsic limitations on entanglement sharing (the
so-called {\em monogamy} constraints \cite{contangle}). We can study
the asymptotic scaling of this entanglement in the limit  $x
\rightarrow \infty$. One finds that the corresponding symplectic
eigenvalue $\eta_{i,i+1}$ is equal to $(N-2)/N$ for even $N$, and
$[(N-2)/N]^{1/2}$ for odd $N$: neighboring sites are thus
considerably
 more entangled if the ring size is even-numbered. Such frustration effect on entanglement
 in odd-sized rings, already devised in a similar context in Ref.~\cite{frusta}, is quite puzzling.
 An explanation may follow from counting arguments
 applied to the number of parameters (which are related to the degree of
 pairwise entanglement) characterizing a generic pure state on
 harmonic lattices \cite{purogen}.

\begin{figure}[t!]
\includegraphics[width=8.5cm]{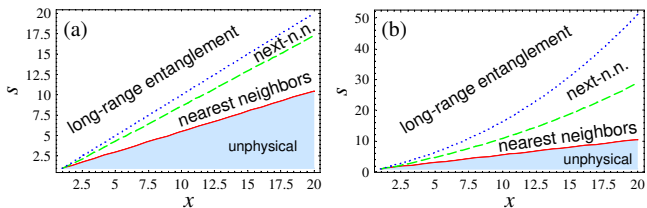}
\caption{(Color online) Entanglement distribution for a six-mode
Gaussian MPS constructed from (a) infinitely entangled EPR bonds
and (b) finitely entangled bonds given by two-mode squeezed states
of the form \eq{tmss} with $r=1.1$. The entanglement thresholds
$s_k$ with $k=1$ (solid red line), $k=2$ (dashed green line) and
$k=3$ (dotted blue line) are depicted as functions of the
parameter $x$ of the building block. For $s > s_k$, all pairs of
sites $i$ and $j$ with $|i-j| \le k$ are entangled (see text for
further details).} \label{thresh}
\end{figure}

\noindent {\em Medium-range correlations.}--- The connection between
input entanglement and output correlation length can be investigated
in detail considering a general $\gr\gamma$ with $s > s_{\min}$. The
MPS CM in \eq{cmout} can still be worked out analytically for a low
number of modes, and numerically for higher $N$. Let us keep the
parameter $x$ fixed; we find that with increasing $s$ the
correlations extend smoothly to distant modes. A series of
thresholds $s_k$ can be found such that for $s > s_k$, two given
modes $i$ and $j$ with $|i-j| \le k$ are entangled. While trivially
$s_1(x) = s_{\min}$ for any $N$ (notice that nearest neighbors are
entangled also for $s=s_1$), the entanglement boundaries for $k>1$
are in general different functions of $x$, depending on the number
of modes. We observe however a
 certain regularity in the process:  $s_k(x,N)$ always increases with the
 integer $k$.
 These considerations follow from analytic calculations on up to
ten-modes MPS, and we can infer them to hold true for higher $N$
as well, given the overall scaling structure of the MPS
construction process. Very remarkably, this means that the maximum
range of bipartite entanglement  between two modes, i.e.~the
maximum distribution of multipartite entanglement, in a Gaussian
MPS on a translationally invariant ring, is {\em monotonically}
related to the amount of entanglement in the reduced two-mode
input port of the building block.
%Moreover, no complete transfer of entanglement to
%more distant modes occurs:  closer sites remain still entangled even
%when correlations between farther pairs arise. This feature will be
%precisely understood in the limit $s \rightarrow \infty$. Before
%that, we present as an example the study of a Gaussian MPS with
%$N=6$ modes.

To clearly demonstrate this intriguing connection, let us consider
the example of a Gaussian MPS with $N=6$ modes. In  a six-site
translationally invariant ring, each mode can be correlated with
another being at most $3$ sites away ($k=1,2,3$). From a generic
building block \eq{bblock}, the $12 \times 12$ CM \eq{cmout} can
be analytically computed as a function of $s$ and $x$. We can
construct the reduced CMs $\gr\gamma^{out}_{i,i+k}$ of two modes
with distance $k$, and evaluate for each $k$ the respective
symplectic eigenvalue $\eta_{i,i+k}$ of the corresponding partial
transpose. The entanglement condition $s > s_k$ will correspond to
the inequality $\eta_{i,i+k} < 1$. With this conditions one finds
that $s_2(x)$ is the only acceptable solution to the equation: $72
s^8 - 12 (x^2 + 1) s^6 + (-34 x^4 + 28 x^2 - 34) s^4 + (x^6 - 5
x^4 - 5 x^2 + 1) s^2 + (x^2 - 1)^2 (x^4 - 6 x^2 + 1)=0$, while for
the next-next-nearest neighbors threshold one has simply
$s_3(x)=x$. This enables us to classify the entanglement
distribution and, more specifically, to observe the interaction
scale in the MPS $\gr\Gamma^{out}$: Fig.~\ref{thresh}(a) clearly
shows how, by increasing initial entanglement in $\gr\gamma_{ss}$,
one can gradually switch on quantum correlations between more and
more distant sites.

\begin{figure}[t!]
\includegraphics[width=8.5cm]{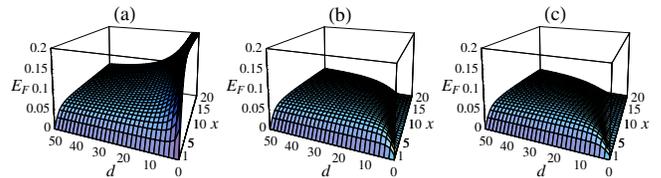}
\caption{(Color online) Entanglement of formation between two
sites $i$ and $j$ in a six-mode Gaussian MPS, with $|i-j|$ equal
to: (a) $1$, (b) $2$, and (c) $3$, as a function of the parameters
$x$ and $d=s-s_{\min}$.} \label{efidi}
\end{figure}

We can also study entanglement quantitatively. Fig.~\ref{efidi}
shows the entanglement of formation $E_F$ of
$\gr\gamma^{out}_{i,i+k}$ for $k=1,2,3$ (being computable in such
symmetric two-mode reductions), as a function of $x$ and $d \equiv
s-s_{\min}$. For any $(x,d)$ the entanglement is a decreasing
function of the integer $k$, i.e.~quite naturally it is always
stronger for closer sites. However, in the limit of high $d$ (or,
equivalently, high $s$), the three surfaces become close to each
other. We want now to deal exactly with this limit, for a generic
number of modes.

\noindent {\em Long-range correlations.}--- In the limit $s
\rightarrow \infty$, the expressions greatly simplify and we
obtain a $N$-mode Gaussian MPS $\gr\Gamma^{out}$ of the form
\eq{circ}, where $C$ and $C^{-1}$ are completely degenerate
circulant matrices, with $(C^{-1})_{i,i}=a_q = [(N - 1) + x^2]/(N
x)$, $(C^{-1})_{i,j\ne i}=c_q =(x^2-1)/(N x)$; and accordingly
$(C)_{i,i}=a_p = [1 + (N - 1)x^2]/(N x)$, $(C^{-1})_{i,j\ne i}=c_p
=-c_q$ . For any $N$, thus, each individual mode is {\em equally
entangled} with any other, no matter how distant they are.

The asymptotic limit of our analysis  shows then that an infinitely
entangled input port of the building block results in a MPS with
{\em maximum} pairwise entanglement length. These $N$-mode Gaussian
states are well-known as useful resources for multiparty CV
communication protocols \cite{network}. The CM $\gr\Gamma^{out}$ of
these MPS can in fact be put, by local symplectic (unitary)
operations, in a standard form parametrized by the single-mode
purity $\mu_{loc} = (a_q a_p)^{-1/2}$ \cite{adescaling}. Remarkably,
in the limit $\mu_{loc} \rightarrow 0$ (i.e.~$x \rightarrow
\infty$), the entropy of any $K$-sized ($K<N$) sub-block of the
ring, quantifying entanglement between $K$ modes and the remaining
$N-K$, is {\em infinite} \cite{adescaling}.
%It is worth reminding that these states, previously known as
%``GHZ-type'' states \cite{network}, have been recently renamed CV
%GHZ/$W$ states, as it has been shown that they exhibit a monogamous
%but promiscuous entanglement sharing \cite{contangle}.
%: in the case e.g. of
%three modes, they possess both maximum tripartite residual
%entanglement and maximum couplewise entanglement between any pair of
%modes.
Within the MPS framework, we also understand the  peculiar
``promiscuous'' entanglement sharing \cite{contangle} of these fully
symmetric states: being them built by a symmetric distribution of
infinite pairwise entanglement among multiple modes, they achieve
maximum genuine multiparty entanglement while keeping the strongest
possible bipartite one in any pair. Let us note that in the field
limit ($N \rightarrow \infty$) each single pair of
 modes is in a separable state, as they have to
mediate a genuine multipartite entanglement distributed among {\em
all} the  infinite modes \cite{adescaling}.

\smallskip \noindent {\em Finitely entangled bonds and experimental
feasibility.}--- We finally discuss how to implement the recipe of
 Fig.~\ref{fiocco} to produce Gaussian MPS experimentally. The three-mode
building blocks can be engineered for any choice of $(x,s)$
(within the practical limitation of a finite available degree of
squeezing) by combining three single-mode squeezed states through
a sequence of up to three beam splitters \cite{3modi}. The EPR
measurements are in turn realized by homodyne detections
\cite{giedke}.

The only unfeasible part of the scheme is constituted by the
ancillary EPR pairs. But are {\em infinitely} entangled bonds truly
necessary? One could consider a $\gr\Gamma^{in}$ given by the direct
sum of two-mode squeezed states of \eq{tmss}, but with finite $r$.
Repeating our analysis to investigate the entanglement properties of
the resulting Gaussian MPS with finitely entangled bonds, we find
that, at fixed $(x,s)$, the entanglement in the various partitions
is degraded as $r$ decreases, as somehow expected. Crucially, this
does not affect the connection between input entanglement and output
correlation length. Numerical investigations show that, while the
thresholds $s_k$ for the onset of entanglement between distant pairs
are quantitatively modified  --  a bigger  $s$  is required at a
given $x$ to compensate the less entangled bonds -- the overall
structure stays untouched. As an example, Fig.~\ref{thresh}(b)
depicts the entanglement distribution  in six-mode MPS obtained from
finitely entangled bonds with $r=1.1$, corresponding to $\approx
6.6$ dB of squeezing. Single-bonded Gaussian MPS, which surprisingly
encompass a broad class of physically relevant multimode states, can
thus be experimentally produced, and their entanglement distribution
can be precisely engineered starting from the parameters of a simple
bisymmetric three-mode building block, with the supply of two-mode
{\em finitely} squeezed states.

\smallskip \noindent {\em Concluding remarks.}--- We have shown that
the  range of pairwise quantum correlations in translationally
invariant $N$-mode Gaussian MPS is determined by the entanglement
in the input port of the building block.
%This interesting
%connection is obtained analytically for any $N$ in the limits of
%minimum and maximum entanglement of the building block, and
%quantitatively analyzed for any intermediate input entanglement in
%an explicit multimode harmonic ring.
%To the best of our knowledge,
%such an interesting connection had not been pointed out  in
%traditional discrete-variable MPS, and further investigation in
%this direction, still resorting to the Jamiolkowski isomorphism,
%may be worthy.
As a consequence of this interesting connection,
%, our findings evidence
a striking difference between finite-dimensional and
infinite-dimensional MPS is unveiled, as the former are by
construction slightly entangled for a low dimensionality of the
bonds \cite{vidal}, and their entanglement is short-ranged
\cite{kore}. We proved instead that pure, fully symmetric,
$N$-mode Gaussian states are exactly MPS with minimum bond
cardinality ($M=1$): yet, their entanglement can {\em diverge}
across any global splitting of the modes, and their pairwise
quantum correlations have {\em maximum} range. How this feature
connects with the potential validity of an area law for critical
bosonic systems  is currently an open question.
%\cite{adescaling}.
%Also the localizable entanglement of a GHZ/$W$
%state, defined as the maximum bipartite entanglement concentrable on
%a single pair of modes by locally performing measurements on all the
%others (and thus an estimator of multipartite entanglement),
%diverges for $x \rightarrow \infty$ \cite{telepoppy}.
%Generalizations to non-Gaussian projectors for the study of
%entanglement  scaling
%in ground states of non-harmonic models, await further investigation.
\\
\noindent We thank I. Cirac,  A. Ekert, F. Illuminati, D. Oi, T.
Osborne, R. Rodriquez, A. Serafini and M. Wolf for valuable
discussions. Financial support from project RESQ (IST-2001-37559) of
the IST-FET programme of the EU is acknowledged. ME is further
supported by The Leverhulme Trust.

\end{document}